\documentclass[12pt,preprint]{aastex}
\usepackage{color}
\usepackage{graphics,graphicx}
\usepackage{textcomp}
\usepackage[varg]{txfonts}
\usepackage{natbib}

\shorttitle{Variations of Dose Rate on Mars}
\shortauthors{Guo et al.}

\begin{document}


\title{Modeling the variations of Dose Rate measured by RAD \\
during the first MSL Martian year: 2012-2014}


\author{
Jingnan Guo\altaffilmark{1},
 Cary Zeitlin\altaffilmark{2},
 Robert~F. Wimmer-Schweingruber\altaffilmark{1}, 
 Scot Rafkin\altaffilmark{3},
 Donald~M. Hassler\altaffilmark{3},
 Arik Posner\altaffilmark{4},
 Bernd Heber\altaffilmark{1},
 Jan K\"ohler\altaffilmark{1},
 Bent Ehresmann\altaffilmark{3}, 
 Jan~K. Appel\altaffilmark{1},
 Eckart B\"ohm\altaffilmark{1}, 
 Stephan B\"ottcher\altaffilmark{1},
 S\"onke Burmeister\altaffilmark{1}, 
 David E. Brinza\altaffilmark{5},
 Henning Lohf\altaffilmark{1}, 
 Cesar Martin\altaffilmark{1},
 H. Kahanp\"a\"a \altaffilmark{6},
 G\"unther Reitz \altaffilmark{7} 
}

\altaffiltext{1}{Institute of Experimental and Applied Physics, Christian-Albrechts-University, Kiel, Germany\email{guo@physik.uni-kiel.de}}\label{inst:kiel}

\altaffiltext{2}{Southwest Research Institute, Earth, Oceans \& Space Department, Durham, NH, USA}\label{inst:durham}

\altaffiltext{3}{Southwest Research Institute, Space Science and Engineering Division, Boulder, USA}\label{inst:boulder}

\altaffiltext{4}{NASA Headquarters, Science Mission Directorate, Washington DC, USA}\label{inst:nasa}

\altaffiltext{5}{Jet Propulsion Laboratory, California Institute of Technology, Pasadena, CA, USA}\label{inst:jpl}

\altaffiltext{6}{Finnish Meteorological Institute, Helsinki, Finland}\label{inst:finn}

\altaffiltext{7}{ Aerospace Medicine, Deutsches Zentrum f\"ur Luft- und Raumfahrt,   K\"oln, Germany}\label{inst:koeln}



\begin{abstract}
The Radiation Assessment Detector (RAD), on board Mars Science Laboratory's (MSL) rover Curiosity, measures the {energy spectra} of both energetic charged and neutral particles along with the radiation dose rate at the surface of Mars. With these first-ever measurements on the Martian surface, RAD observed several effects influencing the galactic cosmic ray (GCR) induced surface radiation dose concurrently: [a] short-term diurnal variations of the Martian atmospheric pressure caused by daily thermal tides, [b] long-term seasonal pressure changes in the Martian atmosphere, and [c] the modulation of the primary GCR flux by the heliospheric magnetic field, which correlates with long-term solar activity and the rotation of the Sun. The RAD surface dose measurements, along with the surface pressure data and the solar modulation factor, are analysed {and fitted to empirical models which quantitatively demonstrate} how the long-term influences ([b] and [c]) are related to the measured dose rates. {Correspondingly we can estimate dose rate and dose equivalents under different solar modulations and different atmospheric conditions, thus allowing empirical predictions of the Martian surface radiation environment. }
\end{abstract}





\section{Introduction and Motivation}\label{sec_intro}

The assessment of the radiation environment is fundamental for planning future human missions to Mars and evaluating the impact of radiation on the preservation of organic bio-signatures.
Contributions to the radiation environment on the Martian surface are very complex \citep[e.g.,][]{saganti2002, dartnell2007modelling, ehresmann2011, koehler2014, ehresmann2014}: 
energetic particles entering the Martian atmosphere either pass through without any interactions with the ambient atomic nuclei, or undergo inelastic interactions with the atmospheric nuclei creating secondary particles (via spallation and fragmentation processes), which may further interact while propagating through the atmosphere.
Finally all primary and secondary particles reaching the surface may also interact with the regolith and, amongst others, produce neutrons which could be backscattered and detected as albedo neutrons \citep[e.g.,][]{boynton2004}.
Therefore, the radiation environment measured at the surface of the planet is determined by the characteristics of the primary radiation incident at the top of the atmosphere, the composition and mass of the atmosphere, and the composition and density of the surface soil. 
The above process can be described by a simplified mathematical equation:
\begin{eqnarray}\label{eq:primary2surface_flux}
F_j(z, t, E) =  \int\limits_{0}^{\infty} \sum_{i} M_{ij} (z, E, E_0) F_{0i}(E_0, t) dE_0,
\end{eqnarray}
where $E_0$ is the energy of a primary particle with species $i$ {(e.g., protons, alpha particles and heavy ions)} reaching the top of the Martian atmosphere; $F_{0i}(E_0, t)$ is the spectrum {(in the unit of counts/MeV/sec/cm$^2$)} of primary particle type $i$ at time $t$; $M_{ij} (z, E, E_0)$ is the yield matrix, representing the interaction between particles and the atmosphere (and the regolith), of particle type $i$ with energy $E_0$ generating particle type $j$ with energy $E$; $M_{ij}$ therefore depends on the altitude $z$ (or atmospheric pressure $P$); finally $F_j(z, t, E)$ {(in the unit of counts/MeV/sec/cm$^2$)} is the resulting particle spectra of type $j$ at time $t$. 
Compared to Neutron Monitors on Earth which measure the count rates of secondary particles generated by primary fluxes going through the atmosphere \citep{clem2000neutron}, RAD measures a mix of primary and secondary particles. Further, there is no need to include the geomagnetic cutoff energy in the case of Mars due to the absence of a global magnetic field.
 
There are predominantly two types of primary particles ($F_{0i}(E_0, t)$) reaching Mars: galactic cosmic rays (GCRs) and solar energetic particles (SEPs). 
SEPs are sporadic and impulsive events and take place much more frequently during the active phase of the solar cycle. 
SEPs are mainly protons, electrons and $\alpha$ particles with energy typically ranging from 10 to several hundreds of MeV. 
GCRs generally originate from outside the Solar System, e.g. in supernova remnants, and their composition consists mainly of protons, $\sim$ 7-10\% helium and $\sim$ 1 \% heavier elements. Due to {the scattering effect of the} magnetic fields in interstellar space, charged GCRs are subject to continuous deflection and the observed spectra are mostly isotropic.
The GCR flux in the Solar System is {inversely} modulated by {the variations of the} solar activity \citep[e.g.,][]{parker1958interaction}. 
{In the long term, d}uring solar activity maximum the increased solar and heliospheric magnetic fields are more efficient at hindering low-energy GCRs from entering the inner heliosphere {\citep[e.g.,][]{heber2007, wibberenz2002simple}} than at solar activity minimum {when the interplanetary magnetic field strength are reduced \citep{goelzer2013analysis, smith2013decline, connick2011interplanetary}}. 
Consequently, the GCR population is most intense during solar minimum {\citep[e.g.,][]{mewaldt2010record, schwadron_lunar_2012}}.
{In the short term, t}he GCR spectrum can also be altered indirectly by solar events {such as coronal mass ejections (CMEs) where} the enhanced interplanetary magnetic field can sweep away a fraction of GCRs causing {reductions in GCR doses in the form of Forbush decreases \citep{forbush1938, schwadron_lunar_2012, schwadron_does_2014}}.

The Mars Science Laboratory (MSL) spacecraft \citep{grotzinger2012mars}, carrying the Curiosity rover, was launched on November 26, 2011 and the rover landed on Mars on August 6, 2012. 
The Radiation Assessment Detector \citep[RAD,][]{hassler2012} onboard is an energetic particle detector and carried out radiation measurements during its cruise from Earth to Mars \citep{zeitlin2013} and now continues to do so on the surface of the planet \citep{hassler2014}. 
The solar modulation of the dose rate measured by RAD during the MSL cruise phase has been discussed by \citet{guo2015a}. 

\begin{figure}
\centerline{\includegraphics[width=1.0\columnwidth]{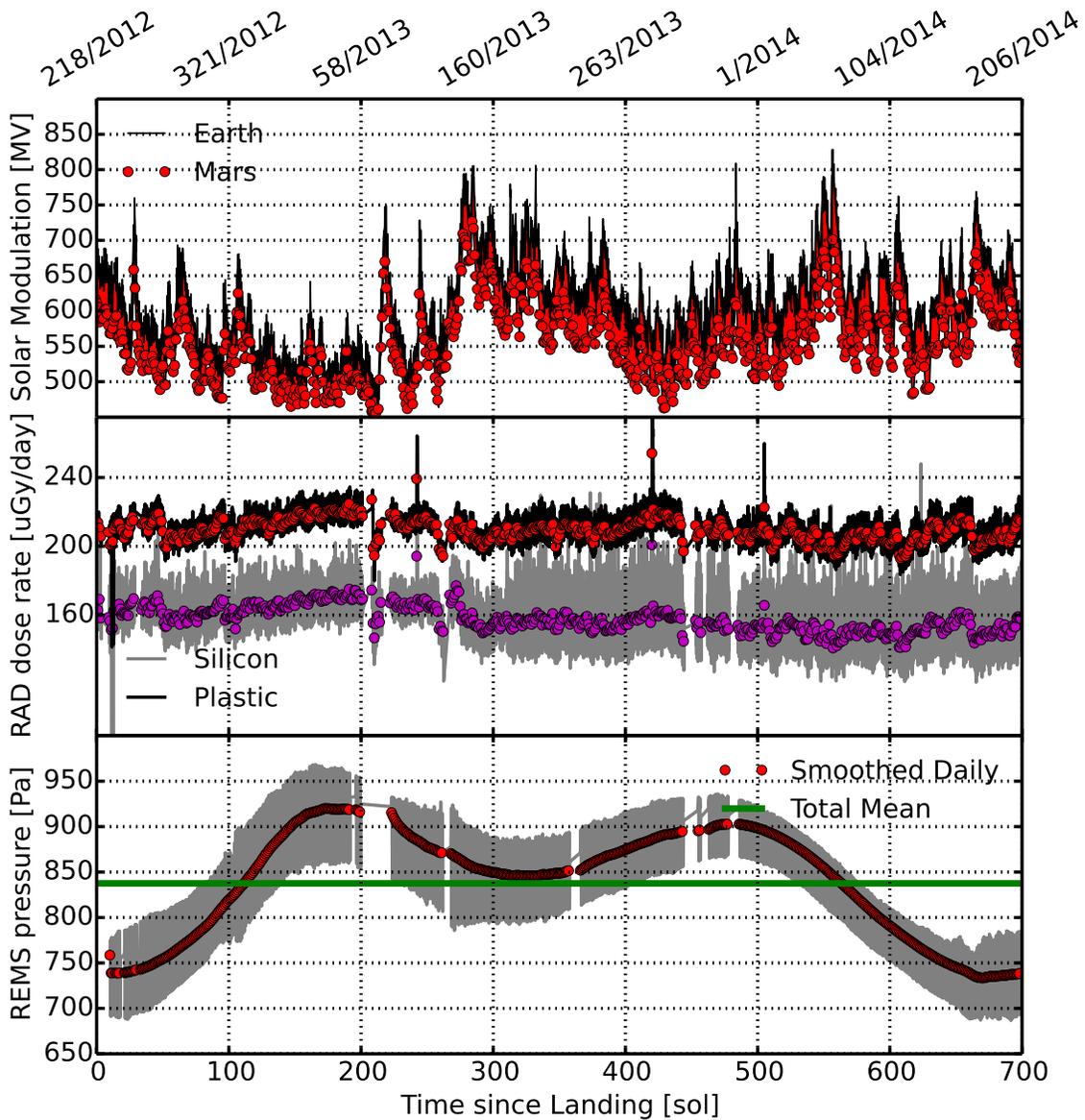}}
\caption{\textit{Top}: Solar modulation potential $\Phi$ at Earth derived from Oulu neutron monitor count rate is shown in black {and the per-sol-averaged $\Phi$ corrected to Mars' location is}  plotted as red dots. 
\textit{Middle}: Dose rates recorded by RAD in the silicon detector B (gray curve) and plastic scintillator E (tissue-equivalent, black curve) with their per-sol-averaged values marked as magenta- and red- dots respectively. 
\textit{Bottom}: The pressure data from REMS (gray curve) and the per-sol-averaged values (red dots). The overall average pressure of the first 700 sols is about 840 Pascal and is marked as a thick-green line. 
{Note that we have given the unit of time in both sol (i.e., time since the landing of MSL) at the bottom of the figure and day of year/year format at the top of the figure.} }\label{fig:surface}
\end{figure}

Typically, CME and SEP events are more common during solar maximum, but the latest solar maximum has been a very weak one compared to space-age averages {\citep[e.g.,][]{schwadron2011coronal, mccomas2013weakest, komitov2013sunspot}. This may have been caused by the reduced solar wind pressure \citep[e.g.,][]{mccomas2008weaker, schwadron2014coronal} in the deep cycle 23-24 minimum which has allowed the termination shock to move closer to the Sun and led to a weakened modulation of the heliosheath \citep{scherer2011cosmic}.}
There were only three solar particle events detected by RAD over its first Martian year \footnote{1 Martian year $\approx$ 668 sols; 1 sol = 1 Mars day $\approx$ 1.03 earth day} on the surface of Mars. They can be seen in the middle panel of Figure \ref{fig:surface} as three peaks of the measured dose rate (the dose rate measurement will be explained in more detail in Section \ref{sec_RAD}) at sols 242, 420, and 504 respectively. The dose rates during these SEP events were several times higher than the quiet-time dose rate. For a closer inspection of the variations of the GCR-driven dose rate during the solar quiet periods, we omit the peak values of the SEPs in this figure. A zoomed-out figure containing the surface dose rate measurements for the first 300 sols, as well as the peak doses of the first SEP can be found in \cite{hassler2014}. 

The bottom panel of Figure \ref{fig:surface} shows the Mars surface atmospheric pressure as measured by REMS \citep{gomez2012rems}, while the middle one shows the dose rate measured by RAD. Both panels show high time resolution data as a shaded band and one-sol averages as solid dots. The variations in the shaded bands are real short-term (diurnal) oscillations in dose rate and are anti-correlated with corresponding diurnal pressure variations. {A zoomed-in figure containing one-sol variations of the dose rate and the pressure can be found in \citet{rafkin2014}}. 
However, some of the long-term evolutions, e.g., the drops in both dose rates and pressure in the time period proceeding sol 200 to those around and after sol 300, can not be well explained by the anti-correlation between them. 
The aim of this paper is to investigate the relative influences of atmospheric pressure (bottom panel) and heliospheric modulation (top panel) on the measured dose rate (middle panel). 

During solar-quiet periods, several factors on different time scales may affect the variation of the GCR-induced dose rates measured by RAD on the surface of Mars: [a] the short-term diurnal variations of the Martian column mass (measured as surface pressure in a hydrostatic atmosphere) at Gale crater caused by daily thermal tides \citep{rafkin2014}, [b] the long-term seasonal changes of the atmospheric pressure 
shown in the bottom panel of Figure \ref{fig:surface}, and [c] the modulation of the primary GCR flux by the solar magnetic field which correlates with the solar activity and the rotation of the heliosphere. A commonly used parameter of heliospheric modulation is the modulation potential $\Phi$ \citep{gleeson1968solar, usoskin2005} which corresponds to the mean electric potential that quantifies the energy loss of a cosmic ray particle experiences inside the heliosphere and is often used to parametrize the modulation of the GCR spectrum. 
{The modulation potential allows specification of distributions across a range of GCR particle species with different nucleons A and charge state Z and  has been often used in determining GCR spectra and flux based on analytic models \citep[e.g.,][]{badhwar2006}.} 
In other words, the GCR-driven primary flux in Equation \ref{eq:primary2surface_flux} is a function of $\Phi$, i.e., $F_{0i}(\Phi)$ and the yield matrix, $M_{ij}$, is a function of atmospheric pressure, $P$; the resulting particle spectra, $F_j(E)$, along with the dose rate which is a good measure of the radiation environment, are consequently a function of $\Phi$ as well as pressure, i.e. $F_j(E) = F_j(\Phi, P, E)$. 
This study aims to derive an empirical expression for the dose rate as a function of $\Phi$ and pressure based on observational data and thus make it possible to predict the radiation environment on the surface of Mars under different solar modulations and pressure variations.
From the neutron monitor count rates ($CR_{NM}$, in the unit of counts per minute) recorded by the Oulu neutron monitor \footnote{The Oulu count rate data have been obtained from http://cosmicrays.oulu.fi/ and the pressure effect has been corrected.}, the modulation potential $\Phi$ (in the unit of MV) at Earth can be estimated \citep{usoskin2002, guo2015a} and its results are plotted in black in the top panel of Figure \ref{fig:surface}.
{However, the potential $\Phi$ at Earth and at Mars may differ due to the longitudinal difference of the modulation across the Parker spirals resulted from three dimensional drifts \citep[e.g.,][]{jokipii2004transverse, potgieter1992simulated} as well as the small radial gradient between 1.0 and 1.5 AU \citep{schwadron2010earth, gieseler2008radial}. This radial effect can be corrected following the analytic function given by \citet{schwadron2010earth} and the resulting modulation $\Phi$ at Mars (per-sol-average) is plotted in red dots. It is generally smaller than $\Phi$ at Earth with variant differences (shown as red-shaded areas) through time due to varying distances of the planets to the Sun.}

Transient effects that are localized to narrow ranges of heliospheric longitude, such as narrow CMEs, can also perturb the GCR fluxes differently at Earth and Mars. 
In order to reduce the spatial longitudinal discrepancy of $\Phi$, we use binning techniques in our current study as presented in Section \ref{sec:Phi_doserates}.

\section{RAD Measurements}\label{sec_RAD}

RAD measures both the charged as well as the neutral radiation environment on Mars \citep{hassler2012}. It uses the ${\rm{d}}E/{\rm{d}}x$ vs total $E$ or multiple ${\rm{d}}E/{\rm{d}}x$ techniques \citep{ehresmann2014, guo2015b} to identify charged particles in a telescope stack of three silicon semiconductor detectors, A, B, and C, followed by a high-density CsI scintillator calorimeter, D. The CsI crystal together with a plastic (BC-432m) scintillator (namely detector E) are enclosed in a highly efficient anticoincidence in order to measure the neutral radiation \citep{koehler2014}. In addition, the dose rate is measured in the silicon detector B as well as in the plastic detector E. A detailed overview of the instrument is given in \citet{hassler2012}.


In this work we will determine the influence of heliospheric modulation and atmospheric pressure on dose rate in silicon (detector B) and in plastic (detector E). The dose rate is defined as the energy deposited by radiation per unit mass and time and is measured in Gy/day (J/kg/day). During quiet times, the dose rate on the surface of Mars is - apart from a very small natural background - determined by the GCR and its interaction with the atmosphere and soil. It can be described by the following equation:
\begin{eqnarray}\label{eq:dose_def}
D(\Phi, P) = \sum\limits_{j} \sum\limits_{area} {\iint \limits_{\epsilon_{min} 0}^{\epsilon_{max} \infty} \lambda_j(E, \epsilon) F_j(\Phi, P, E) dE d\epsilon}/{m}, 
\end{eqnarray}
where $E$ is the energy of a particle with type $j$, $F_j(\Phi, P, E)$ {(in the unit of counts/MeV/sec/cm$^2$)} is the surface particle spectrum (equivalent to $F_j(z, t, E)$ in Equation \ref{eq:primary2surface_flux}) which is modulated by the heliospheric potential, $\Phi$, and atmospheric pressure, $P$. $m$ {(kg)} is the mass of the detector and $\epsilon$ is the energy deposited by the particle in the detector. This energy deposit can be estimated using either a simple Bethe-Bloch Ansatz \citep{bethe1932bremsformel} or with more sophisticated {models such as GEANT4 \citep{agostinelli2003} and HZETRN \citep{wilson1995hzetrn}} and is included as a yield matrix, $\lambda_j(E,\epsilon)$, in the above equation. $\epsilon_{min}$ and $\epsilon_{max}$ are the minimum and maximum energy over which the detector is sensitive and $D$ is the corresponding dose rate {integrated over the entire detection area ($area$) and all the detected particle species. Dose rate is in the unit of MeV/kg/sec and can be transfered to \textmu Gy/day}. Correspondingly $D$ depends on both heliospheric potential, $\Phi$, and on atmospheric pressure, $P$. 

RAD measures radiation doses induced by both charged and neutral energetic particles in two detectors: the silicon detector B and the plastic scintillator E \citep{hassler2012}. 
{RAD is directly mounted on the 'shoulder' of the rover body \citep{hassler2012} and the shielding of the rover from above can be ignored. It is indeed shielded by the rover from below. However since the upward flux (of albedo particles) is much smaller than the downward flux \footnote{More detailed analysis shows that the upward-downward ratio of the proton flux is only about 13\% (Appel et al., paper in preparation).}, the measured dose can be roughly assumed equivalent to the radiation dose of the Martian surface. }

RAD operates on adjustable "observation" cycles, with typical durations of 32 minutes early in the surface mission and 16 minutes later in the mission to date.
These cadences typically yield 44 or 88 measurements per day or sol for dose rates or charged/neutral particle count rates.
Detector E has a composition similar to that of human tissue and is also more sensitive to neutrons than silicon detectors. 
For a given incident flux, the dose rate in detector B is generally less than the dose rate in E because of the comparatively larger ionization potential of silicon as shown in Figure \ref{fig:surface}.

\section{Martian Atmospheric Pressure}
The Martian atmosphere is roughly 1\% as thick as that of the Earth's. The dominant composition of the Martian atmosphere is about 95\% CO$_2$ \citep{owen1977composition}, of which 25\% condenses seasonally onto the winter pole. Nevertheless, the seasonal variation of the composition has a very little effect on the surface radiation field compared to the changes of the column mass \citep[e.g.,][]{rafkin2014}. The oscillations in atmospheric column mass drive variation of energetic particle radiation at the surface with both diurnal and seasonal periods.
 
\subsection{Diurnal Pressure Variations} \label{sec:dirunal_pressure}

With a constant value of gravitational acceleration $g$, the surface pressure is an exact measure of the column mass given a hydrostatic atmosphere. An increase in column mass corresponds to an increase of surface pressure \citep[e.g.,][]{rafkin2014}.

The Martian atmosphere exhibits a strong thermal tide excited by direct solar heating of the atmosphere on the dayside and strong infrared cooling on the nightside. Heating causes an inflation of the atmosphere with a simultaneous drop in surface pressure.
In Gale Crater, the thermal tide produces a diurnal variation of column mass of about $\pm 5\%$ relative to the median, as measured by the Rover Environmental Monitoring Station (REMS) \citep{gomez2012rems}. 
The magnitude of the diurnal pressure cycle at Gale Crater is substantially greater than previous surface measurements. This is likely due to the topography of the crater environment, which yields hydrostatic adjustment flows that amplify the daily tides \citep{haberle2014preliminary}.
 
This daily pressure oscillation can be seen in the bottom panel of Figure~\ref{fig:surface} where the pressure (shown as gray lines) expands away from the daily averaged pressure (shown as red dots) within the range of $\approx \pm$ 50 Pascals.

\subsection{Seasonal Pressure Changes}\label{sec:seasonal_pressure}

The seasonal Martian atmospheric pressure variation is controlled by a complex balance between the cold and warm poles \citep[e.g.,][]{tillman1988mars, zurek1988}. 
As on Earth, when the south pole is in total darkness, the north pole is experiencing continuous sunlight; one might expect that the global pressure should stay roughly constant over the year, as CO$_2$ vaporized at one pole would freeze at the other. However, the high eccentricity of Mars' orbit causes the insolation to be significantly different between poles.  
Mars is farther from the Sun during northern summer; the summer in the southern hemisphere is much warmer than summer in the northern hemisphere. 
As a result, the north and south poles have different impacts on the atmospheric pressure changes through condensation of CO$_2$ to the polar region in winter and the recession of CO$_2$ polar cap during spring and summer.    
The seasonal CO$_2$ condensation cycle results in the seasonal pressure variation: as more CO$_2$ evaporates into the atmosphere in the summer, the measured surface pressure increases as shown in the bottom panel of Figure~\ref{fig:surface}. 

Curiosity landed at the time when the northern hemisphere was in late summer and the global pressure was near its minimum since the southern CO$_2$ ice cap had nearly reached its maximal extent during southern hemisphere winter. 
As shown in Figure \ref{fig:surface}, the atmospheric pressure then began to increase as the southern polar cap started shrinking during the southern spring (northern autumn). It then reached a peak ($\sim$ sol 175) during early northern hemisphere winter, when the southern cap was near its minimum level and before the northern cap had grown to its maximum size. A small minimum of the pressure is present during the late northern winter when the northern cap reaches its maximum extent ($\sim$ sol 310).

The diurnal pressure oscillation (as described in Section \ref{sec:dirunal_pressure}) superimposed on this long term seasonal pressure change can be averaged out by calculating the per-sol-averaged pressure as shown by the red dots in Figure \ref{fig:surface}. The peak to peak seasonal pressure difference reaches about 25\% of the average pressure. 
This seasonal pressure variation, like the diurnal variation, causes column density changes of the atmosphere that  affect the particle fluxes measured by RAD at a seasonal period.  

\section{Empirically Modeling the Dose Rate as a Function of Pressure and Solar Potential}

As discussed in Section \ref{sec_intro}, we aim to find out how RAD measured surface radiation environment depends on atmospheric pressure $P$ and solar modulation $\Phi$. 
Both influences are blended and embedded in the long-term variations of the measured GCR dose rate as seen in Figure \ref{fig:surface}. 
While some previous studies have attempted to model the effects of atmospheric pressure on the Martian radiation environment and dose rate \footnote{\citet{ehresmann2011} calculated the surface radiation exposure for much higher atmospheric pressures which might have been present during the Noachian epoch.}, this work has the advantage of using actual in-situ measured data for a quantitative and empirical study.

\citet{rafkin2014} have analyzed the diurnal pressure effect on the dose rate measurement by successfully isolating the diurnal variations in the RAD measurements from the longer-term influences that include seasonal atmospheric shielding and variability of the heliosphere. This method will be explained in Section \ref{sec:diurnal_fit_mtd}. 
The authors presented a robust fit with a linear correlation between pressure changes and dose rate measurements \footnote{This linear correlation has been obtained at the scale of pressure values measured at Gale Crater and should not be simply extrapolated over a much wider range of pressure, e.g., to the top ot the Martian atmosphere.} providing the reasonable assumption that the solar modulation $\Phi$ varies at a time scale longer than the diurnal period. 
\citet{guo2015a} have studied the solar modulation of the GCR dose rate measured by RAD during the MSL cruise phase when the pressure-variation effect was not present. 
Two separate empirical models were employed to describe the anti-correlation between heliospheric modulation potential $\Phi$ and measured dose rate.  
Both a simple linear function and a non-linear regression model could equally well represent this anti-correlation given that the shielding of the spacecraft ('pressure') was constant.
Assuming that $\Phi$ and pressure $P$ are two independent parameters influencing the surface dose rate (while the viability of this assumption will be discussed in Section \ref{sec_discuss}), the pressure effect and the solar modulation can be linearly combined. This results into the following two models of GCR-induced dose rate variations on the surface of Mars.

\begin{enumerate}
\item Both the pressure effect and the solar modulation drive the variation of dose rate linearly and independently, written as:
\begin{eqnarray}\label{eq:model1_pres_phi}
D(\Phi, p) = D_{01} + \kappa P + \beta \Phi,
\end{eqnarray}
where $\kappa$ {(in the unit of \textmu Gy/day/Pa)} is the linear correlation factor between pressure and dose rate variations and can be fitted when $\Phi$ is constant; $\beta$ {(in the unit of \textmu Gy/day/MV)} is the linear correlation factor between solar potential and dose rate changes and can also be fitted when pressure is approximately stable; $D_{01}$ {(in the unit of \textmu Gy/day)} is some relevant dose rate when both pressure and $\Phi$ are at certain typical levels, e.g. $P_0$ and $\Phi_0$.

\item The solar modulation effect can also be described by a non-linear empirical model \citep{guo2015a} as often used in the analysis of neutron count rates \citep[e.g.,][]{usoskin2011}. The combination of the pressure and $\Phi$ changes results in:  
\begin{eqnarray}\label{eq:model2_pres_phi}
D(\Phi, P) = D_{02} + \kappa P + \frac{\alpha_1}{\Phi+ \alpha_2} ,
\end{eqnarray}
with the requirement of $\alpha_2 \geq 0 $ to assure a positive denominator. 
$\alpha_1$ {(in the unit of $\rm{MV \cdot Gy} /day$)} and $\alpha_2$ {(in the unit of MV)} can be fitted when pressure is constant.
$D_{02}$ {(in the unit of \textmu Gy/day)} is some relevant dose rate when both pressure and $\Phi$ are at certain typical levels.

\end{enumerate}    

\subsection{Pressure Effect}\label{sec:pressure_effect}
The day and night variations of the dose rate have been observed for the first time on the surface of Mars by RAD, as presented by \citet{rafkin2014} where the anti-correlation between pressure and dose rate changes has been quantitatively investigated.
Due to the day and night oscillations of the atmospheric column mass, the characteristics of the particle radiation at the surface also vary diurnally. 
The middle panel in Figure \ref{fig:surface} shows the dose rate measured by RAD on the surface of Mars during the first 700 sols (06/Aug/2012 - 25/Jul/2014). 
The dose rate varies at a diurnal level seen as oscillations of the black curve for the plastic detector, E, and the gray curve for the silicon detector, B. The variation in the E dose rate over a diurnal cycle averages about 15 \textmu Gy/day (or $\sim$5 \%)  peak to peak, out of around 220 \textmu Gy/day. 
The oscillation in the silicon detector is more pronounced because it includes both the diurnal variation and larger statistical fluctuations   due to its smaller geometric factor. The per-sol-averaged values of the dose rates are calculated and plotted over the curves as magenta-dots (for B measurements) and red-dots (for E measurements). 
Three SEP events occurred on sol 243, 421 and 504 and they are excluded from the following study of the pressure effect since the particle fluxes and energies during SEPs are substantially different from those of GCRs and the pressure response of dose rate can be heavily modified.
Quantitative analysis has shown a clear inverse relation between the variations in the atmospheric pressure and the RAD dose rates with a correlation coefficient for linear regression of 96\%. 

\subsubsection{Fitting $\kappa$ using Hourly Perturbation} \label{sec:diurnal_fit_mtd}
We use the method described in \citet{rafkin2014} to produce the average diurnal perturbations of the data; this approach aims at isolating the diurnal variations in the RAD measurements from the longer-term influences that include seasonal atmospheric shielding and variability of the heliosphere, as well as solar event.

Generally, the dose rates measured by RAD are distributed uniformly in time while REMS' pressure data are recorded at 1 Hz over 5 minute periods at certain periods of the sol. In order to correlate these two data sets with different time frames, we obtain the hourly averaged measurements of both pressure and dose rates and their corresponding hourly perturbations.
The hourly averaged dose rate and pressure are $D_{h,s}$ and $P_{h,s}$ respectively where $h$ represents the hour (here defined to be 24 hours on each sol) and $s$ corresponds to the sols. 
The measurements, especially the pressure data, were not always uniformly taken in time and there are gaps in the data lasting longer than one hour. Therefore we only consider sols where there are at least 20 hours with pressure measurements in each hour. Then we interpolate the pressure data using a spline interpolation method so that data are distributed uniformly and continuously over 24 hours through that sol and systematic errors of the hourly average and per-sol-average could be minimized. Finally the interpolated pressure data can be binned into 24 bins for each sol and the hourly binned pressure is the corresponding hourly average $P_{h,s}$. 
The hourly pressure perturbation $\delta P_{h,s}$ is defined as the difference between the hourly average, $P_{h,s}$, and the total average, $P_{s}$, of the corresponding sol: 
\begin{eqnarray}\label{eq:hourly_pert}
\delta P_{h,s} = P_{h,s} - P_{s} = P_{h,s} - \frac{\sum\limits_{h=1}^{24} P_{h,s}}{24}.
\end{eqnarray}
In this case, the hourly perturbation is isolated from the sol-to-sol changes; $\delta P_{h,s}$ can further be averaged through all sols to obtain the mean hourly perturbation $\bar{\delta P_{h}}$ and the corresponding standard error in that hour.
The same binning technique has also been applied to RAD dose rate measurement.
The mean hourly perturbation of dose rate, $\bar{\delta D_{h}}$, can be readily correlated with the hourly pressure perturbation, $\bar{\delta P_{h}}$, and their relationship follows a clear anti-correlation which can be fitted with a first-order polynomial function:
\begin{eqnarray}\label{eq:hourly_pert_correlation}
\bar{\delta D_{h}} = \kappa_{d} \times \bar{\delta P_{h}}.
\end{eqnarray}

\subsubsection{Pressure Effect on Dose Rates}\label{sec:pres_dose}
\begin{figure}
\centerline{\includegraphics[width=\columnwidth]{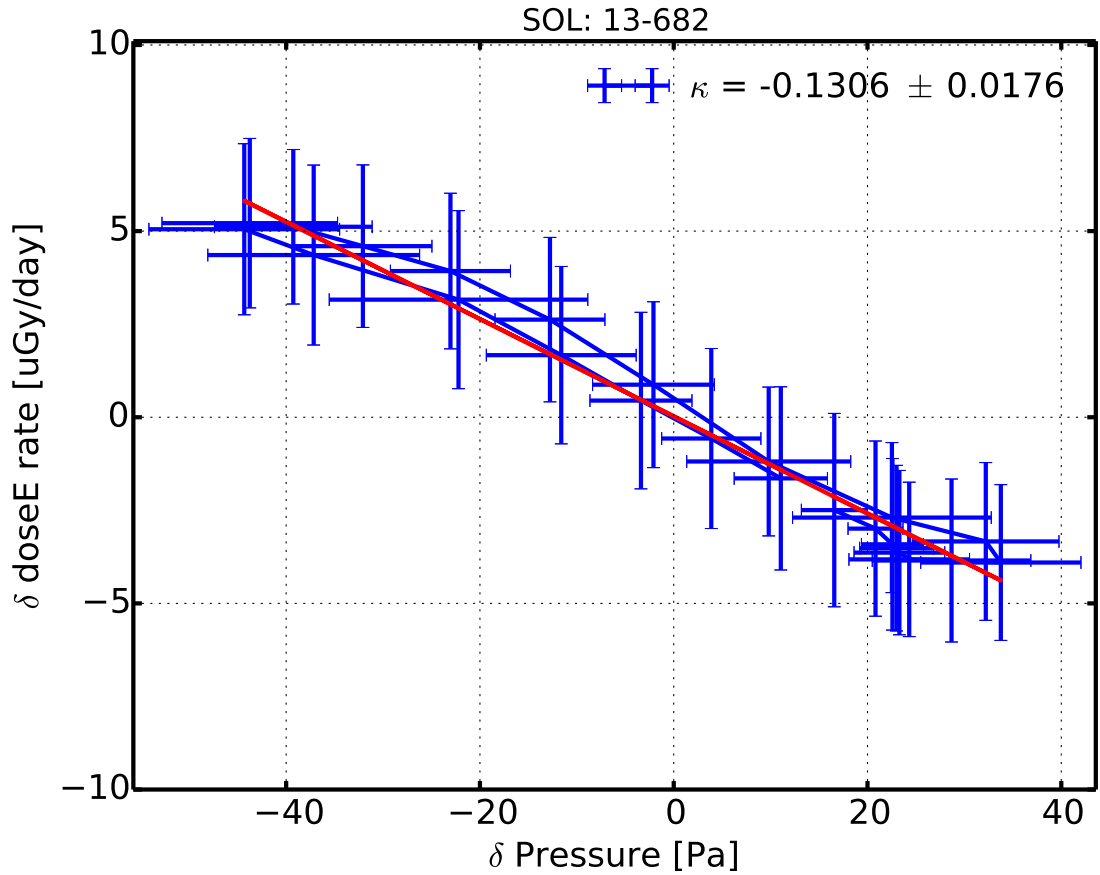}}
\caption{Hourly perturbation of dose rate $\bar{\delta D_{h}}$ versus hourly perturbation of pressure $\bar{\delta P_{h}}$ through sol 13 to 682 (approximately one Martian year) as shown in blue. The error bars stand for the standard deviation of the averaged hourly perturbation.
The fitted anti-correlation is shown as a red line with a slope of $\kappa_d$ being -0.1306 $\pm$ 0.0176 \textmu Gy/day/Pa.} 
\label{fig:hourlyPert_doseE}
\end{figure}

Using the binning method described in Section \ref{sec:diurnal_fit_mtd}, we fitted the diurnal dose rate variations (measured by detector E) as a response of pressure oscillations through the first MSL Martian year as shown in Figure \ref{fig:hourlyPert_doseE}. The correlation coefficient for linear regression is 98.8\%. 
The fitted proportionality factor, $\kappa$, is -0.13 $\pm$ 0.02 \textmu Gy/day/Pa where the error bar is obtained through propagation of the uncertainties of the hourly perturbation data. The measured dose rate decreases when the pressure increases since $\kappa$ is negative. This indicates that the atmosphere has a shielding effect on surface dose rates. The proportionality factor obtained here is consistent, within error bars, with what obtained by \citet{rafkin2014} which was about 0.116 \textmu Gy/day/Pa. The slight difference is caused by the selection of different periods (\citet{rafkin2014} used the first 350 sols of data), and by the method used here to interpolate data into a uniform distribution in time before the binning process. 
We have applied the same method to fit the pressure effect on dose rates deposited in detector B and found the correlation coefficient to be 96.4\% and the proportionality factor to be -0.10 $\pm$ 0.06 \textmu Gy/day/Pa which is smaller than that for the E detector since dose rates in B are generally smaller.

Note that compositional changes in the atmosphere have only a vanishingly small effect on surface radiation compared to the pressure variations \citep{rafkin2014}. Similarly, the occasional presence of dust in the atmosphere during dust storms is negligible\footnote{A PLANETOCOSMIC simulation has been carried out to derive the particle fluxes on the surface of Mars considering the existence of dust storms. The result shows that the effect of dust storms is very small (Appel et al., paper in preparation).}. 

\subsection{Solar Modulation}\label{sec:solar_modulation}
The pressure variations and the modulation of the primary GCR radiation outside the atmosphere are the two factors which determine the longterm dose rate variations on the surface of Mars as explained in Section \ref{sec_RAD}. In order to separate these two effects, we {subtract} the pressure effect using the proportionality of the variations found above. 


\subsubsection{Subtracting the Pressure Effect} \label{sec:subtractP}
\begin{figure}
\centerline{\includegraphics[width=1.0\columnwidth]{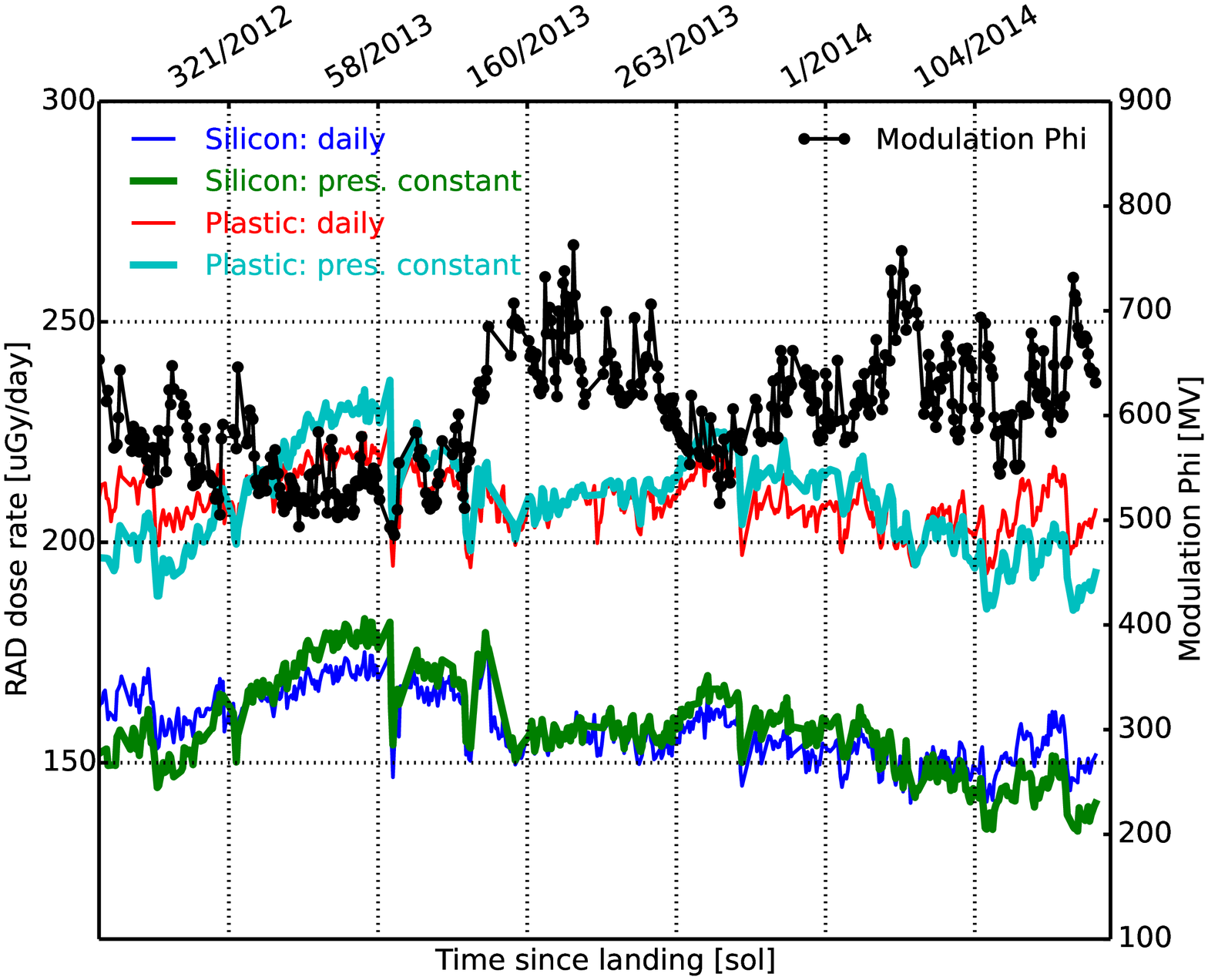}}
\caption{RAD dose rates (\textmu Gy/day, left axis) and solar modulation potential $\Phi$ {at Mars' radial distance} (MV, think-black-dotted-line, right axis) through sol 13 to 682 (approximately one Martian year). 
RAD dose rates from silicon detector B is in blue and from plastic detector E is in red.
Thick-green and thick-cyan lines are the resulting dose rates ($D'$ in Equation \ref{eq:model_pres_subtract}) assuming pressure is constant at $P_0$. 
{Note that we have given the unit of time in both sol (i.e., time since the landing of MSL) at the bottom of the figure and day of year/year format at the top of the figure.} }
\label{fig_pressure_correction}
\end{figure}      
As we can already isolate the pressure effect and obtain the proportionality factor $\kappa$ in Equation \ref{eq:model1_pres_phi} and \ref{eq:model2_pres_phi}, the effect of solar modulation can be investigated by first subtracting the pressure effect from dose rate and then correlating the 'constant-pressure' dose rate with solar modulation potential $\Phi$.  
The method can be described by restructuring, for instance Equation \ref{eq:model1_pres_phi} as following:
\begin{eqnarray}\label{eq:model_pres_subtract}
D' (\Phi) = D_{01}' + \beta \Phi,
\end{eqnarray}
where $D'= D - \kappa(P-P_0) $ and $ D_{01}'= D_{01} + \kappa P_0$.
Employing all the per-sol-averaged dose rate data collected over one Martian year (sol 13 to 682), we can subtract the seasonal pressure effect assuming the surface pressure is constant at a particular value, i.e., $P_0=$ 840 Pascals, which is the averaged pressure found by REMS during this time period as shown in Figure \ref{fig:surface}.
By calculating the difference of the per-sol-averaged pressure $P$ and $P_0$ and using $\kappa$ found in Section \ref{sec:pres_dose}, we estimate the pressured-induced dose rate to be $\kappa(P-P_0)$ which is then subtracted from the long-term dose rate measurement with the 'constant-pressure-as-$P_0$' dose rate remaining, namely $D'$ being solely a function of solar modulation potential $\Phi$.
Figure \ref{fig_pressure_correction} shows the RAD dose rates measured by both silicon and plastic detectors before (thin lines) and after (thick lines) the pressure correction. Also plotted is the solar modulation $\Phi$ which already shows an anti-correlation with the constant-pressure dose rates.

\subsubsection{Correlation of dose rates and $\Phi$} \label{sec:Phi_doserates}
\begin{figure}
\centerline{\includegraphics[width=1.0\columnwidth]{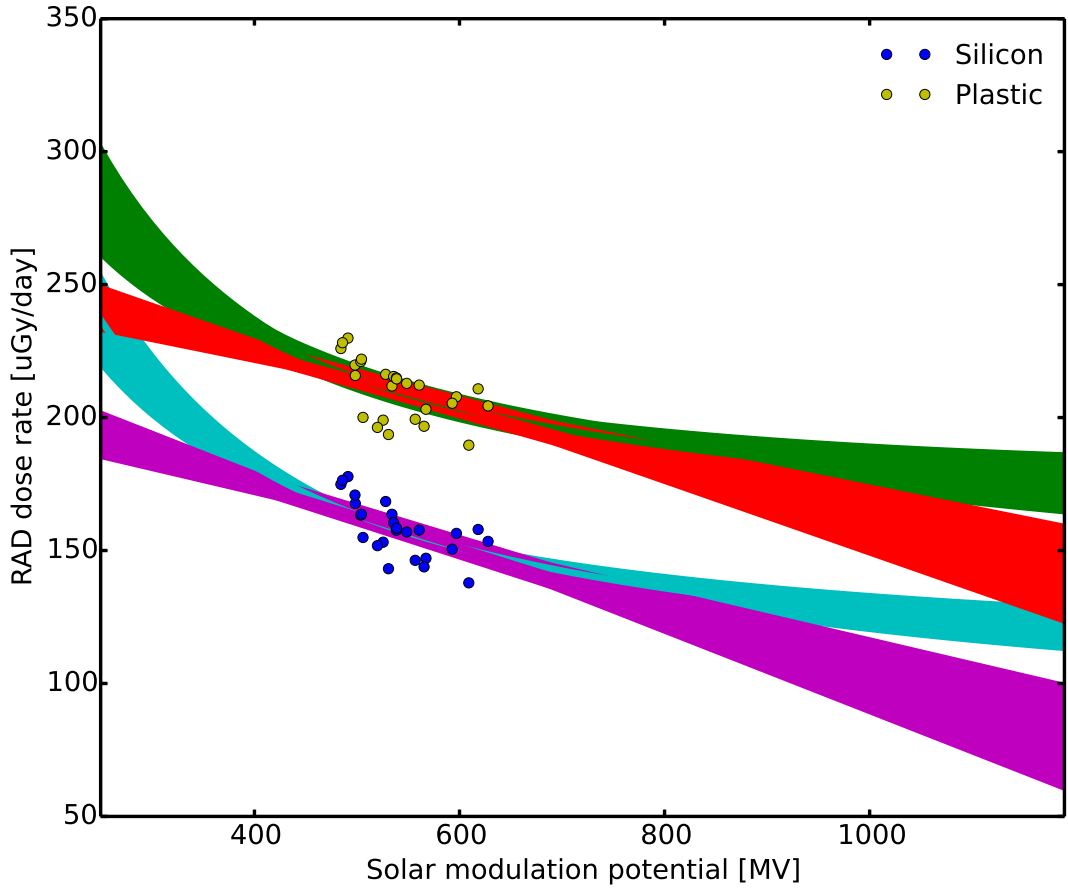}}
\caption{26-sol binned data and fittings processed through bootstrap Monte Carlo simulations of the variations of the RAD dose rate (with pressure assumed to be constant $P_0=840$ Pa) and the solar modulation $\Phi$ on the surface of Mars through sol 13 to 681.
The red/magenta line and area represent the linear fits with standard errors (Eq.~\ref{eq:model_pres_subtract}) of constant-pressure dose rate in detector E/B versus solar modulation potential. The green/cyan line and area show the results of the non-linear fits (Eq.~\ref{eq:model2_pres_phi}). } 
\label{fig_dose_Phi_fits}
\end{figure}

The modulation potential, $\Phi$, is derived from the Oulu neutron monitor on Earth {(at 1AU) and corrected to the radial distance of Mars ($\sim$ 1.5 AU) following \citet{schwadron2010earth}. However} Mars and Earth are not always magnetically well connected {and cross-field diffusion and drift can be extremely important. In other words, the modulation process is fundamentally 3-dimensional and} $\Phi$ can not directly represent modulations at Mars\footnote{During the cruise phase MSL was mostly magnetically connected with the Earth \citep{posner2013, guo2015a} and directly correlating $\Phi$ measured at Earth and the RAD dose rate was sensible.}.
In order to average out the cross-field discrepancy of the modulation between Earth and Mars and to smooth out the rotation of the heliospheric magnetic fields, we bin the 
per-sol-averaged data as shown in Figure \ref{fig_pressure_correction} into 26-sol-averaged bins and correlate the binned $\Phi$ values and constant-pressure dose rates.
The correlation coefficients are {-0.66} and -0.56 for detectors B and E respectively, clearly indicating a negative correlation between solar modulation and GCR induced dose rates.
We applied regression fittings of both the linear and non-linear models described by Equations \ref{eq:model1_pres_phi} and \ref{eq:model2_pres_phi} to data from both detectors as shown in Figure \ref{fig_dose_Phi_fits}.

In order to reliably propagate the uncertainties due to both measurements and binning processes, we have carried out a bootstrap Monte Carlo simulation \citep{efron1981nonparametric}.
Simulated data sets are generated using the uncertainty range of the binned data which are then fitted to the models; dose rate results at a wider range of $\Phi$ are estimated simultaneously at each fit; 500 simulated fits were processed for each model; for every $\Phi$ value ranging from 250 to 1200 MV, 500 dose rate values were generated and their mean is taken to be the 'predicted' dose rate while their standard deviation as the uncertainty. 
The fitting results of the two separate models applied to both detectors are presented below. 
%
%
%
%

\begin{itemize}

\item The fitted parameters $\beta$ and $D_{01}'$ for the linear model are obtained as the mean values of the 500 fitted parameters and their uncertainties are the standard deviation of the 500 Monte Carlo fits. 
For the silicon detector, B, we obtained {the parameters of the linear model as $D_{01}' = 224.10 \pm 16.12 $ \textmu Gy/day and $\beta = -0.12 \pm 0.03$ \textmu Gy/day/MV.}
For plastic detector E,{ the fitted parameters are $D_{01}' = 267.17 \pm 16.29 $ \textmu Gy/day and $\beta = -0.11 \pm 0.03$ \textmu Gy/day/MV.}
Note that the absolute values of $D_{01}'$ are not essential to our study and they depend on the choice of normalized pressure, $P_0$. 
The parameter $\beta$ however, as shown in Equation \ref{eq:model_pres_subtract}, directly represents the linear dependence of dose rate changes on the solar modulation potential, $\Phi$. 
The results obtained from measurements of the two different detectors are consistent with each other within error bars.     


\item The non-linear model has three fitting parameters: $\alpha_1$, $\alpha_2$ and $D_{02}' = D_{02} + \kappa P_0$. Their values and error bars are also obtained using the same Monte Carlo method. 
{For silicon detector, B, the fitted parameters are $D_{02}' = 89.4 \pm 14.0 $ MV, $\alpha_1 = (3.7 \pm 0.8) \times 10^4$ MV \textmu Gy/day and $\alpha_2 = (9.9 \pm 0.1) \times 10^{-4}$ MV. 
For plastic detector, E, the results are $D_{02}' = 149.6 \pm 15.3 $ MV, $\alpha_1 = (3.3 \pm 0.8) \times 10^4$ MV \textmu Gy/day  and $\alpha_2 = (1.3 \pm 0.1) \times 10^{-3}$ MV. }
The small absolute values of $\alpha_2$ indicates that it could be ignored in the model for the range of data in the current measurement.
\end{itemize}

The 'predicted' dose rates (with constant surface pressure $P_0$) at given $\Phi$ values from 250 MV to 1200 MV were estimated for both models and are plotted in Figure \ref{fig_dose_Phi_fits}. 
The uncertainty of the predicted dose rate increases when the extrapolation is further away from the actual measurements. The linear model often predicts a smaller dose rate than the nonlinear model and this difference is bigger for small solar potentials, i.e., during solar minimum. For instance, at $\Phi = 250$ MV the discrepancy between the two models is as large as 80 \textmu Gy/day for the silicon detector, B.   
Because the predictions of the two models differ substantially at solar extreme conditions, choosing one or the other for predicting the radiation exposure of an astronaut would make a big difference. The data themselves do not rule out one or the other because the time period for which they are currently available does not cover a sufficiently large range of $\Phi$. This underlines the importance of acquiring more data over an extended period of time to adequately cover the entire solar activity cycle. 

\subsubsection{Estimates of the Surface Radiation Environment under different $\Phi$ and $P$}

\begin{figure}
\centerline{\includegraphics[width=1.0\columnwidth]{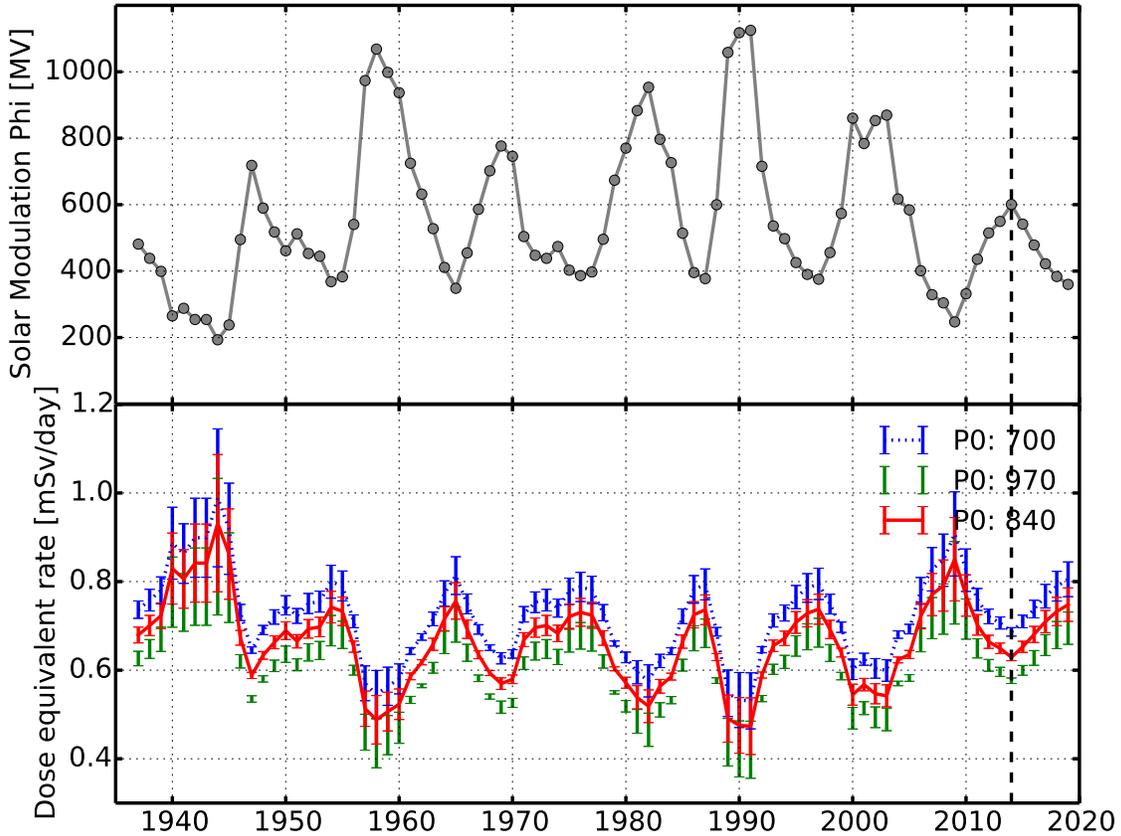}}
\caption{\textit{Top Panel:} Annual average of reconstructed modulation potential $\Phi$ {at the radial distance of Mars} (gray dotted line) since 1937 until 2014 (on the left side of the vertical dashed line) as well as the predicted values from year 2015 to 2019 (on the right side of the vertical dashed line). \textit{Bottom Panel:} Modeled annual dose equivalent rate following the evolution of $\Phi$ assuming different seasonal surface pressures: blue for $p_0 = 700$ Pa, green for $p_0 = 970$ Pa and red for $P_0=840$ Pa. } 
\label{fig_dose_Phi_extrapolates}
\end{figure}

The annually averaged modulation potential $\Phi$ reconstructed from Oulu neutron monitor data \citep{usoskin2011} from 1937 until 2014 can be used to calculate the corresponding 'expected' dose rate predicted by our models. 
We can also use the sunspot number predicted by the U.S. Dept. of Commerce, NOAA, Space Weather Prediction Center (SWPC) for the coming years (2015 to 2019) to estimate the corresponding solar potential following a correlation study of monthly $\Phi$ and sunspot numbers \citep{guo2015a}. 
The annual modulation potential, $\Phi$, {extrapolated to 1.5 AU radial distance from the Sun} through year 1937 until 2019 is shown in the top panel of Figure~\ref{fig_dose_Phi_extrapolates}. 
It shows a clear 11-year cycle and varies from less than 250 MV to more than 1200 MV. 

For evaluating the space radiation environment, dose equivalent is often derived and can be assumed to be proportional to the risk of lifetime cancer induction via population studies (ICRP60). 
Its approximated value, in Sieverts (Sv), is taken to be ,$\rm{<Q>}$ $\times$ $\rm{D}$ , where $\rm{D}$ is measured tissue-equivalent dose and $\rm{<Q>}$ is the average quality factor which is a conventional parameter for radiation risk estimation. 
A $\rm{<Q>}$ value of 3.05 $\pm$ 0.26 was found from RAD's measurements on the surface of Mars over the first 350 sols \citep{hassler2014}.
Multiplying the tissue-equivalent dose rates (directly measured by the plastic detector E or modeled by e.g. HZETRN model) with the average quality factor yields the dose equivalent rate of GCR fluxes.
For dose rates measured by detector B, a silicon to water conversion factor of 1.38 has to be applied first \citep{zeitlin2013} which approximately relates energy loss per unit of path length (${\rm{d}}E/{\rm{d}}x$) in silicon to Linear Energy Transfer (LET) in water.
{RAD measured dose equivalent is thus most comparable to high-water content skin dose equivalent. The body effective dose can be further derived as the weighted sum of different organ dose equivalent (skin, eye, bone, brain, heart, etc.) and the weighting factor for each organ can be found in the National Council on Radiation Protection and Measurements \citep{linton2003national}.}  

The extrapolated dose rates at different modulation potential values have been estimated via two different models and two different detectors as shown in Figure \ref{fig_dose_Phi_fits}. 
These four sets of modeled dose rates can be transfered into dose equivalent rate and the results mostly agree with each other within error bars \footnote{Figure 4 in \citet{guo2015a} has shown the modeled dose equivalent rate during the cruise phase estimated by both linear and non-linear models based on dose rates from both detectors. The results are consistent within error bars with exceptions during extreme solar conditions for which the non-linear model predicts higher dose rates.}.  
Because we can not currently determine which of the two models is the better approximation when the solar potential is outside the measured range, we let all the four modeled values serve as possible results and take their mean as predictions and the propagated errors as uncertainties. The thus final modeled dose equivalent rates and their standard deviations are shown in Figure \ref{fig_dose_Phi_extrapolates}.   

Note that the modeled dose equivalent rate has been derived based on the 'constant-pressure-at-840 Pa' dose rate (see Section \ref{sec:subtractP}). 
However, the seasonal surface pressure at Gale crater, as shown in Figure \ref{fig:surface}, expands between 700 and nearly 1000 Pascals which would result in roughly 0.1 mSv/day of dose equivalent rate changes \footnote{250 Pascals of pressure change would lead to 32.5 \textmu Gy/day of dose rate difference which is then transfered to dose equivalent rate via the quality factor.} and this is a considerable portion $\sim 16 \%$ of annual averaged surface dose equivalent rate $\sim 0.7$ mSV/day. 
In order to show this seasonal pressure effect, we also estimated the dose equivalent rate (at various solar potentials) when pressure is 700 and 970 Pa respectively as shown Figure \ref{fig_dose_Phi_extrapolates}.         
The dose equivalent rates are inversely related to the surface pressure, although the seasonal pressure influence is much smaller than the longer term effects driven by solar modulation. 

The estimated surface dose equivalent rate ranges from about 0.35 mSv/day to about 1.15 mSv/day and has a clear anti-correlation with $\Phi$ as expected from both models.
Stronger solar modulation leads to a decrease of dose equivalent rate and at $\Phi >$ 1000 MV the dose equivalent rate can be as low as $\sim$ 0.35 mSv/day within the uncertainties, considerably smaller than the current averaged measurement. 
At solar maximum and minimum when the modulation potential is further away from the measured range, the uncertainty of the estimations increases due to the large discrepancy between the models. Future measurements over solar minimum periods are essential for improving the predictions at low modulation potentials.

{Due to the shielding of the atmosphere, the current surface dose equivalent rate is only about 40\% of the RAD cruise measurement $\sim$ 1.8 mSv/day \citep{zeitlin2013} and that from the Cosmic Ray Telescope for the Effects of Radiation (CRaTER) on the Lunar Reconnaissance Orbiter $\sim$ 1.6 mSv/day \citep{schwadron_does_2014}. Consequently, our estimations of the surface dose equivalent rate over solar minimum and maximum periods are much smaller than the predictions based on deep space measurements given by RAD \citep[cruise phase,][]{guo2015a} and CRaTER \citep{schwadron_does_2014}. The predictions of dose equivalent rates behind these spacecraft shielding conditions are between about 1 mSv/day (solar maximum) and 5 mSv/day (solar minimum). }

\section{Discussion and Conclusions}\label{sec_discuss}
We have presented the dose rate data collected by MSL/RAD on the surface of Mars and analyzed its short-term and long-term variations which are driven by the atmospheric pressure changes and solar modulation.
Following \citet{rafkin2014}, we first analyzed the dose rate dependence on diurnal pressure oscillations over the first MSL Martian year. 
This pressure-driven effect on dose rate changes in the longterm is then subtracted to recover the constant-pressure dose rate which is then correlated with solar modulation potential $\Phi$.  
A clear anti-correlation, with a correlation coefficient {of about -0.6} between $\Phi$ and the recovered dose rate, suggests that, as expected for higher solar activity, GCR particles are more attenuated and the dose rate is decreased. 

We carried out a quantitative study of this anti-correlation and fitted two models to the measured data using a bootstrap Monte-Carlo method to estimate the uncertainties. The predictions of the two models for solar activity minimum differ substantially and the data are insufficient to decide which of the two models should be used. {\citet{schwadron_does_2014} have estimated the deep space dose rate at different modulation potential derived from HZETRN model and the dose rate is indeed non-linearly dependent on $\Phi$. However the shape (parameters) of this analytically derived model is different from that of our empirically modeled function.} This highlights the need for extended measurements to cover solar activity minimum and an entire activity cycle {and these observations can be used to constrain the analytic models}.

The extrapolated dose equivalent rate at various modulation potentials and different surface pressures are shown in Figure \ref{fig_dose_Phi_extrapolates}. 
The predicted dose equivalent rate during solar maximum years (e.g year 1991) when $\Phi \sim 1100$ MV was found to be as low as $\sim$ 0.35 mSv/day, which is considerably lower than the current surface measurement $\sim$ 0.7 mSv/day since the current solar maximum is atypically quiet. 
The modeled dose equivalent rate under solar minimum conditions can be as high as 1.15 mSv/day within the uncertainties.
The seasonal pressure changes may affect the estimated dose equivalent rate at a level of about 0.1 mSv/day. Although this is less than the long-term solar modulation effect, it should not be ignored. 
Based on the solar modulation potential predicted for the next five years \citep{guo2015a}, we estimate a trend of increasing dose equivalent rate (between 0.56 and 0.84 mSv/day) from 2015 until 2020. 

The correlation between heliospheric potential and RAD-measured dose rates during the cruise phase was investigated by \citet{guo2015a} with the data modeled by both linear and non-linear functions. 
The linear dependence $\beta$ of dose rate on $\Phi$ was -0.39 $\pm$ 0.07 \textmu Gy/day/MV for the silicon detector B and -0.44 $\pm$ 0.04 \textmu Gy/day/MV for the plastic detector E. 
In comparison, $\beta$ derived here for the surface case are much smaller: -0.12 $\pm$ 0.03 and -0.11 $\pm$ 0.03 \textmu Gy/day/MV for B and E respectively. 
This is because [a] the magnetic connection between solar modulation potential $\Phi$ measured at Earth and dose rates evaluated during the cruise was very good, allowing the direct and thus stronger correlation of the daily values; and more importantly: [b] low energy particles which are more affected by solar modulation make a bigger contribution to the GCR doses detected during the cruise phase than to the doses measured at the surface of Mars.
The process of a GCR spectrum penetrating through the Martian atmosphere can be simulated using for instance the PLANETOCOSMICS toolkit \citep{desorgher2006planetocosmics}. 
Given typical Martian atmospheric conditions, we have found that protons with energies less than 170 MeV do not reach the bottom of Gale Crater. 
The penetration energy is species dependent and increases with increasing ion charge.
Instead, the spacecraft shielding was highly non-uniform while nearly 50\% of incoming particle trajectories within the field of view of RAD from above were only lightly shielded \citep{zeitlin2013}, thus allowing more low-energy particles to contribute to the dose rate. 
We will carry out more GEANT4 \citep{agostinelli2003} and PLANETOCOSMICS simulations {as well HZETRN modelling in the future to study the interactions of GCR spectra with different shielding conditions (spacecraft and the atmosphere) in order to derive modeled function of the dose rate dependence on pressure and solar potential which can be compared with our empirical functions.}   

Both models have assumed the independent effect of pressure and $\Phi$ on dose rates. However, this may be modified when pressure and $\Phi$ change over wider ranges than have been observed to date: 
\begin{itemize}
\item a much thinner atmosphere will allow more lower-energy particles to reach the surface which experience stronger modulation (e.g., bigger $|\beta|$ in the linear model). Therefore, for significant pressure changes, $\beta$ could be a function of $P$ , i.e., $\beta(P)$; 

\item much stronger solar modulation (bigger $\Phi$) would lead to a larger fraction of high-energy particles in the GCR flux and these energetic particles are less affected by the atmosphere (smaller $|\kappa|$); when pressure is much higher and the surface atmospheric depth is approaching the Pfotzer maximum \citep[e.g.,][]{richter2008}, most primary particles are shielded while more secondary particles are generated and this may result in a decreased shielding effect; therefore the dependence of dose rate on pressure may be modified as $P$ and $\Phi$ change substantially, i.e., $\kappa = \kappa(P, \Phi)$;            

\item both the linear and non-linear models are empirical and derived from measurements; despite the robustness of the fitting of the actual data, the extrapolation is highly uncertain and a complete model requires measurements over the full range of solar conditions.
\end{itemize}
{To verify these non-linear and second-order effects of our fitted parameters in the empirical function, Monte Carlo simulations as well as analytic HZETRN modeling will be carried out for constraining the parameters and comparing the predictions.}

The quality factor used to derive equivalent dose from dose may also be sensitive to substantial changes in solar modulation because the estimation of the average quality factor depends on the spectra of particles depositing energies in the detector \citep{schwadron_does_2014}.
Hitherto, the measured $\rm{<Q>}$ has been quite stable since [a] the RAD measurements have undergone only small changes of the solar modulation and [b] the modulation affects more the low-energy ions which are more likely to be shielded by the atmosphere and contribute little to the surface spectra. 
The change of the quality factor under much different solar conditions needs to be investigated with extended measurements.

A total Mars mission GCR dose equivalent can be estimated based on our measurements and predictions of both the surface case and the cruise phase \citep{guo2015a}.
The fastest round trip with on-orbit staging and existing propulsion technologies has been estimated to be a 195-day trip (120 days out, 75 days back with an extra e.g., 14 days on the surface), as described by \citet{folta2012fast}. During solar maximum periods when $\Phi \sim 1200 MV$, this would result in a GCR-induced cruise dose equivalent of $195 \pm 98$ mSv and surface dose equivalent of $4.9 \pm 2.0$ mSv, which adds to $200 \pm 100$ mSv during the total mission.

Additional contributions to dose rate and dose equivalent rate by SEPs should not be ignored and they can differ significantly from the current measurements due to the high variability of their frequencies and intensities. 
Small, "soft-spectrum" solar events where particle energies are modest will have little or no effect on the surface dose due to atmospheric shielding. 
For instance the total dose equivalent from the first SEP event observed on Mars (on sol 242) was only 0.025 mSv \citep{hassler2014} while the cruise SEP events had dose equivalent ranging from 1.2 mSv/event to 19.5 mSv/event \citep{zeitlin2013}.  
Future measurements of much bigger SEP events on the surface of Mars are crucial for understanding extreme conditions of radiation environment on Mars for potential manned missions.

%

%

%
%
%
%
%

\acknowledgments
RAD is supported by the National Aeronautics and Space Administration (NASA, HEOMD) under Jet Propulsion Laboratory (JPL) subcontract \#1273039 to Southwest Research Institute and in Germany by DLR and DLR's Space Administration grant numbers 50QM0501 and 50QM1201 to the Christian Albrechts University, Kiel. Part of this research was carried out at JPL, California Institute of Technology, under a contract with NASA. 
The sunspot data has been obtained from: WDC-SILSO, Royal Observatory of Belgium, Brussels. We are grateful to the Cosmic Ray Station of the University of Oulu and Sodankyla Geophysical Observatory for sharing their Neutron Monitor count rate data. 
The data used in this paper are archived in the NASA Planetary Data System’s Planetary Plasma Interactions Node at the University of California, Los Angeles. The archival volume includes the full binary raw data files, detailed descriptions of the structures therein, and higher-level data products in
human-readable form. The PPI node is hosted at \url{http://ppi.pds.nasa.gov/}.

\bibliographystyle{apj}
\bibliography{msl_rad_guo}

\end{document}